\documentclass[twoside]{article}
\usepackage{listings}
\usepackage{xcolor}
\usepackage{graphicx}
\usepackage{multicol}
\usepackage{natbib}
\usepackage{rotating}
\usepackage{hyperref}
\usepackage[sc]{mathpazo} 
\usepackage[T1]{fontenc} 
\usepackage{microtype} 
\usepackage{booktabs} 
\usepackage[english]{babel} 
\usepackage[hmarginratio=1:1,top=32mm,columnsep=20pt]{geometry}
\usepackage{enumitem}
\usepackage{listings}
\usepackage{color}
\definecolor{codegreen}{rgb}{0,0.6,0}
\definecolor{codegray}{rgb}{0.5,0.5,0.5}
\definecolor{codepurple}{rgb}{0.58,0,0.82}
\definecolor{backcolour}{rgb}{0.95,0.95,0.92}
 
\lstdefinestyle{mystyle}{
    backgroundcolor=\color{backcolour},   
    commentstyle=\color{codegreen},
    keywordstyle=\color{magenta},
    numberstyle=\tiny\color{codegray},
    stringstyle=\color{codepurple},
    basicstyle=\footnotesize,
    breakatwhitespace=false,         
    breaklines=true,                 
    captionpos=b,                    
    keepspaces=true,                 
    numbers=left,                    
    numbersep=5pt,                  
    showspaces=false,                
    showstringspaces=false,
    showtabs=false,                  
    tabsize=2
}
 
\usepackage{abstract} 

\usepackage{titlesec} 
\renewcommand\thesection{\Roman{section}} 
\renewcommand\thesubsection{\roman{subsection}} 
\titleformat{\section}[block]{\large\scshape\centering}{\thesection.}{1em}{} 
\titleformat{\subsection}[block]{\large}{\thesubsection.}{1em}{} 

\usepackage{fancyhdr} 
\usepackage{titling} 

\usepackage{hyperref} 

\def\abba{\textsc{A}\reflectbox{\textsc{b}}\textsc{bA}}
\def\mh{$\mu$Hz}

\pretitle{\begin{center}\Huge\bfseries} 
\posttitle{\end{center}} 
\title{Release note: Massive peak bagging of red giants in the \textit{Kepler} field} 
\author{%
\textsc{T. Kallinger} \\[1ex] 
\normalsize Institut f\"ur Astrophysik, Universit\"at Wien \\ 
\normalsize \href{mailto:thomas.kallinger@univie.ac.at}{thomas.kallinger@univie.ac.at} 
}
\date{\today} 
\renewcommand{\maketitlehookd}{%
\begin{abstract}
\noindent The NASA satellite Kepler has gathered about 1\,420 days-long photometric time series for more than 20\,000 red giant stars. For about 6\,600 of them also APOGEE spectroscopic parameters are available, making the sample of high interest for various astrophysical investigations. To optimally exploit the full wealth of the seismic information, extraction of mode parameters of all significant individual frequencies is necessary.  However, the complex structure of the mixed mode pattern makes it challenging to automate the peak bagging (i.e., the extraction of the individual mode parameters from the stars power density spectra). Even though several approaches have been successfully implemented, the available results are still limited to a handful of stars. Here I present frequencies, amplitudes, and lifetimes of more than a quarter of a million  oscillation modes of the spherical degree $l=0$ to 3, which have been observed in 6\,179 Kepler red giants. The sample covers evolutionary stages from the lower red-giant branch to high up the asymptotic giant branch. The modes were extracted with the \underline{A}utomated \underline{B}ayesian Peak-\underline{B}agging \underline{A}lgorithm (\abba ) and are publicly available at \url{https://github.com/tkallinger/KeplerRGpeakbagging}.
\end{abstract}
}

\begin{document}

\maketitle

\section{Introduction}
All cool stars with a convective surface layer show high-overtone oscillations of low spherical degree $l$. These so-called solar-type oscillations are intrinsically damped and stochastically excited by the turbulent flux of the near-surface convection. To characterise them, one typically fits Lorentzians,
\begin{equation}
PDS(\nu) = \frac{a^2\tau}{1+4[\pi\tau(\nu-\nu_c)]^2}
\end{equation}
 to the power density spectrum (PDS) of the observations with the central frequency $\nu_c$, the height (or alternatively the rms amplitude $a$, i.e., the square root of the area under the profile), and the lifetime $\tau$ as free parameters.

\begin{figure}[]
	\begin{center}
	\includegraphics[width=\textwidth]{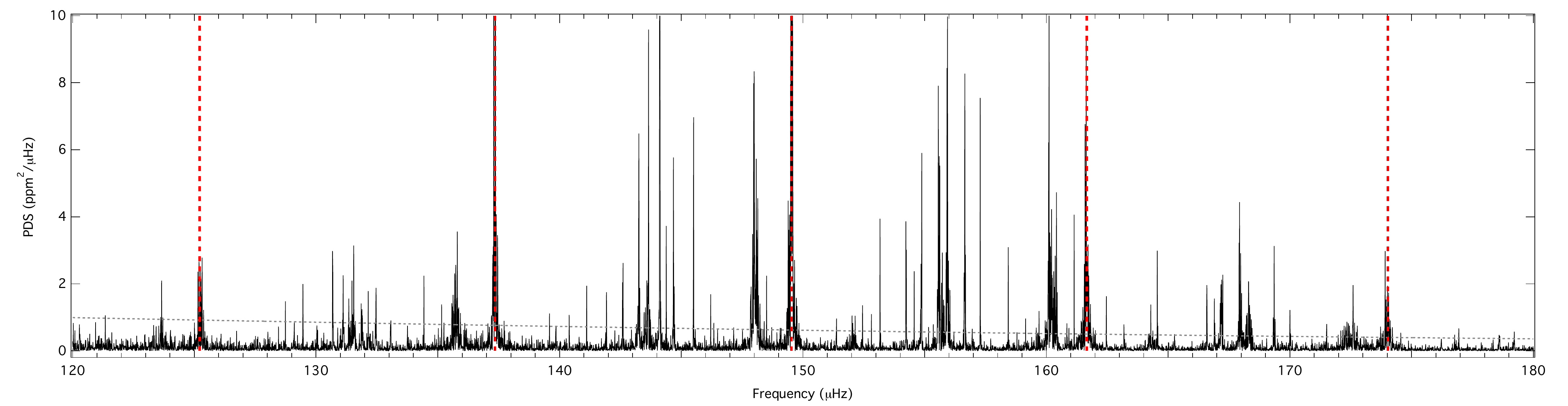}
	\caption{Power density spectrum of the central four radial orders of KIC\,1433803. Radial modes are indicated by red vertical dashed lines to guide the eye. The dotted black line corresponds to eights times the granulation background as sort of significance limit for the indiviual modes.} 
	\label{fig:rawspec} 
	\end{center} 
\end{figure}

For more  than 20\,000 red giants in the original \textit{Kepler} field, detailed “peak bagging” -- meaning the extraction of individual mode parameter -- has only been performed for a handful of stars \citep[e.g.][]{Corsaro2015,GarciaOrtiz2018,Themessl2018,Handberg2016}. This is mainly due to complications introduced by the complex pattern of the many non-radial mixed p/g modes \citep[e.g.][]{Beck2011} in between the regular radial p-modes. The PDS of a single p-mode radial order of a red giant can easily include more than ten observable mixed $l=1$ (dipole) modes. They are often further split into duplets or triplets by rotation \citep[e.g.][]{Beck2012}, so that a given spectrum may  consist of more than 100 individual modes with 300+ (more or less) independent parameters (see Fig.\,\ref{fig:rawspec}). This is further complicated by the fact that the lifetime (i.e., the inverse width) of dipole modes is strongly modulated with the modes relative position in the p-mode radial order, so that modes closer to the centre of the radial order have relatively short lifetimes, while modes closer to the borders live much longer. In the about 1\,420 days-long \textit{Kepler} observations it often appears that the central dipole modes are well resolved (and are therefore to be fitted with a Lorentzian), while the outer modes are unresolved and need to be fit with a sinc$^2$ function,
\begin{equation}
PDS(\nu) = \frac{2a^2}{\delta\nu_\mathrm{bin}}\mathrm{sinc}^2[\frac{\pi(\nu-\nu_c)}{\delta\nu_\mathrm{bin}}],
\end{equation}
where $\delta\nu_\mathrm{bin}$ is the width of independent frequency bins in the PDS. Understandably, fitting more than 100 individual mode profiles, where it is a-priori unknown what functional form they should have, is a quite complex task. Most of the currently existing peak-bagging methods require substantial human intervention. Even though such methods were very successful for individual stars \citep[e.g.][]{DiMauro2018,Beck2018,Huber2019}, they are not scalable to the large number of stars so that the scientific potential of the large amount of data delivered by \textit{Kepler} has not yet been fully exploited. The situation will become even more demanding, with the growing number of incoming data from the TESS and forthcoming Plato mission.

Therefore, a reliable peak-bagging algorithm that is largely automated and free from significant human input would greatly benefit the analysis of large samples of stars. To deliver consistent results, such algorithm would need to automatically detect all observable modes in a given PDS, identify if they are resolved or not, extract the mode parameters and uncertainties, and label their spherical degree, based on a minimum of a-priori assumptions. It further needs to perform this in a reasonable amount of computation time, which rules out fitting all modes at once (with up to hundreds of free parameters) and even performing the fit per radial order can be computationally demanding. 

Here, I present results from the \underline{A}utomated \underline{B}ayesian Peak-\underline{B}agging \underline{A}lgorithm (\abba) that fulfil all of the above mentioned criteria and which delivers a full set of mode parameters in typically less than 25 minutes of computation time (for a PDS based on a 1\,420 days-long \textit{Kepler} time series) on a standard desktop computer. The results are available at \url{https://github.com/tkallinger/KeplerRGpeakbagging}.

\begin{figure}[]
	\begin{center}
	\includegraphics[width=0.5\textwidth]{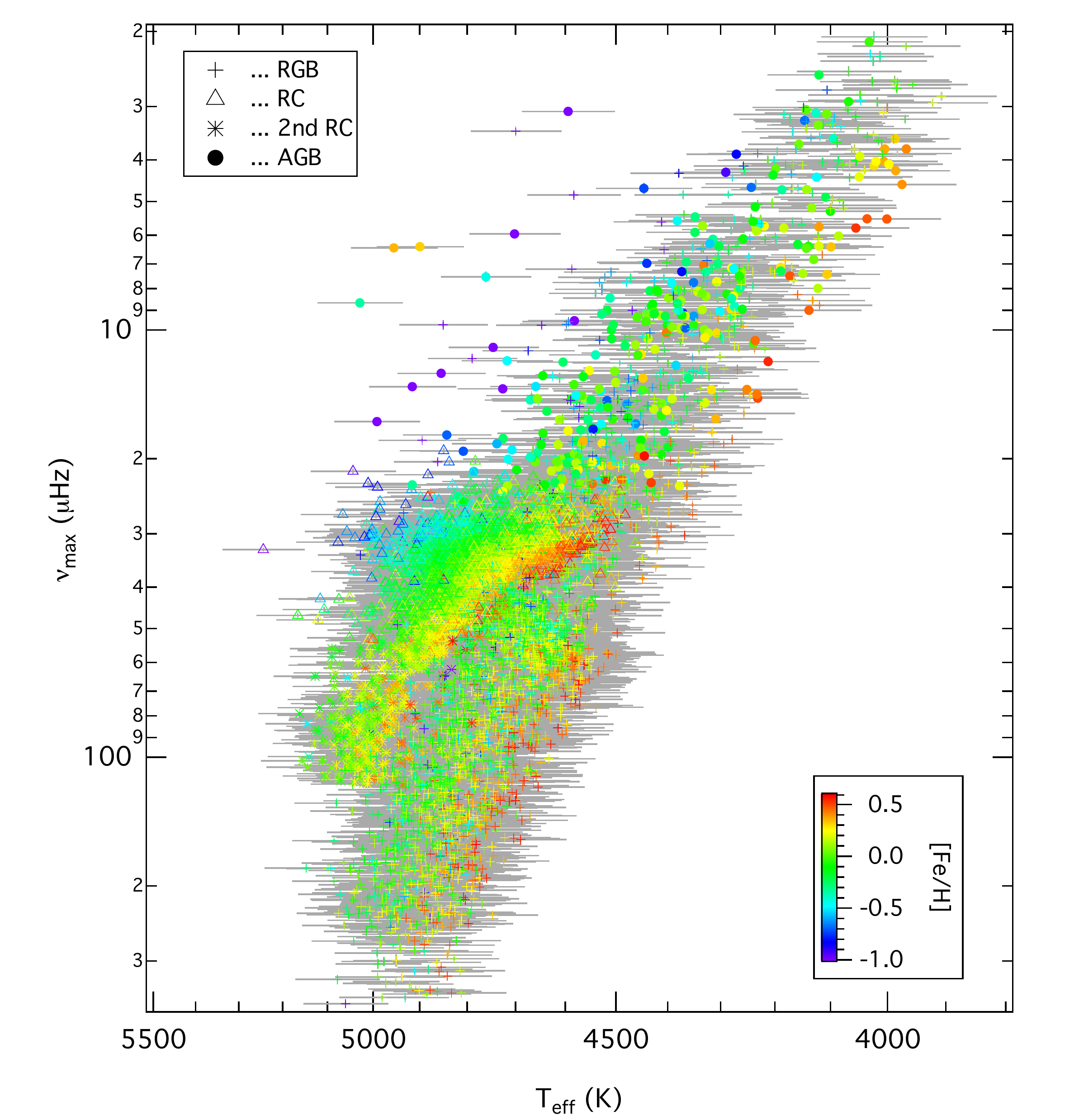}
	\caption{Seismic HR-diagram showing the full sample of APOKASC red giants with the metallicity color-coded. The different symbols indicate the evolutionary stage.} 
	\label{fig:HRD} 
	\end{center} 
\end{figure}

\section{Data sample}
The results presented here were computed for stars that are part of the so-called APOKASC sample, which combines timeseries observations from the \textit{Kepler} mission with near-infrared spectroscopy from the APOGEE Survey for more than 6\,600 red giants \citep{Pinsonneault2018}. The full sample is shown in Fig.\,\ref{fig:HRD}, from which \abba\ has extracted oscillation mode parameters in at least three radial orders for 6\,179 stars covering evolutionary stages from the lower red-giant branch to high up the asymptotic giant branch. For all of these stars spectroscopic temperature and metallicity estimates are available in addition to a total of more than 250\,000 individual oscillation modes making the sample a powerful test-bench for stellar astrophysics. For the remaining 421 stars of the sample, \abba\ failed to deliver some results because the frequencies or their signal-to-noise level are too low, or the modes are reflected by the Nyquist frequency leading to an inverted pattern.
A summary of the number of extracted modes sorted by spherical degree of the modes and evolutionary stage of the star is shown in Tab.\,1.
\begin{table}[b]
\caption{Number of stars in the various evolutionary stages and the extracted modes of different spherical degrees $l$.}
\centering
\begin{tabular}{lrrrrr|r}
\toprule
evo.&\# of stars & $l=0$ & $l=1$ & $l=2$ & $l=3$ &$\sum$\\
\midrule
RGB &3\,375& 24\,138 & 89\,162 & 21\,592 & 5\,493& 140\,385\\
RC &2\,121& 14\,490 &65\,882  & 12\,967& 1\,353& 94\,692\\
2ndRC &448& 3\,776 & 13\,063 & 3\,578& 440& 20\,857\\
AGB &235& 1\,092 & 642 & 1\,113& 0 & 2\,847\\
\midrule
$\sum$& 6\,179 & 43\,496 & 168\,749 & 39\,250 & 7\,286& 258\,781\\
\bottomrule
\end{tabular}
\end{table}

\section{Prerequisites to run AbbA}
Prior \abba\ can extract the mode parameters from the power density spectrum of a given star, the granulation background in the vicinity of the oscillation power excess needs to be characterised. This is done following the approach of \cite{Kallinger2014} and model the global shape of the PDS with the superposition of two super-Lorentzian functions and a Gaussian, where the latter serves as a proxy for the oscillation power with the central frequency representing the frequency of maximum oscillation power ($\nu_\mathrm{max}$). In a next step, a sequence of Lorentzian functions is fit on top of the previously determined background. The sequence covers the $l =0$ to 2 modes in the central three radial orders around $\nu_\mathrm{max}$, with the mode frequencies being parameterised by the frequency of the central radial mode ($\nu_\mathrm{c}$) and some frequency separations \citep{Kallinger2010b}. The resulting large and small separations $\Delta\nu_\mathrm{c}$ and $\delta\nu_{02}$, respectively, represent a local value in the centre of the power excess. The evolutionary stage of the star is determined from the phase shift of the central radial mode \citep{Kallinger2012}. 

Based on the granulation background, the evolutionary stage, and estimates for $\nu_\mathrm{c}$, $\Delta\nu_\mathrm{c}$, and $\delta\nu_{02}$, \abba\ is able to automatically extract the mode frequency, amplitude, and lifetime (including realistic uncertainties) of all modes present in a given PDS, to rate the statistical significance of these modes, and to label their spherical degree. 

\section{The \underline{A}utomated \underline{B}ayesian Peak-\underline{B}agging \underline{A}lgorithm -- AbbA}
\begin{figure}[t]
	\begin{center}
	\includegraphics[width=\textwidth]{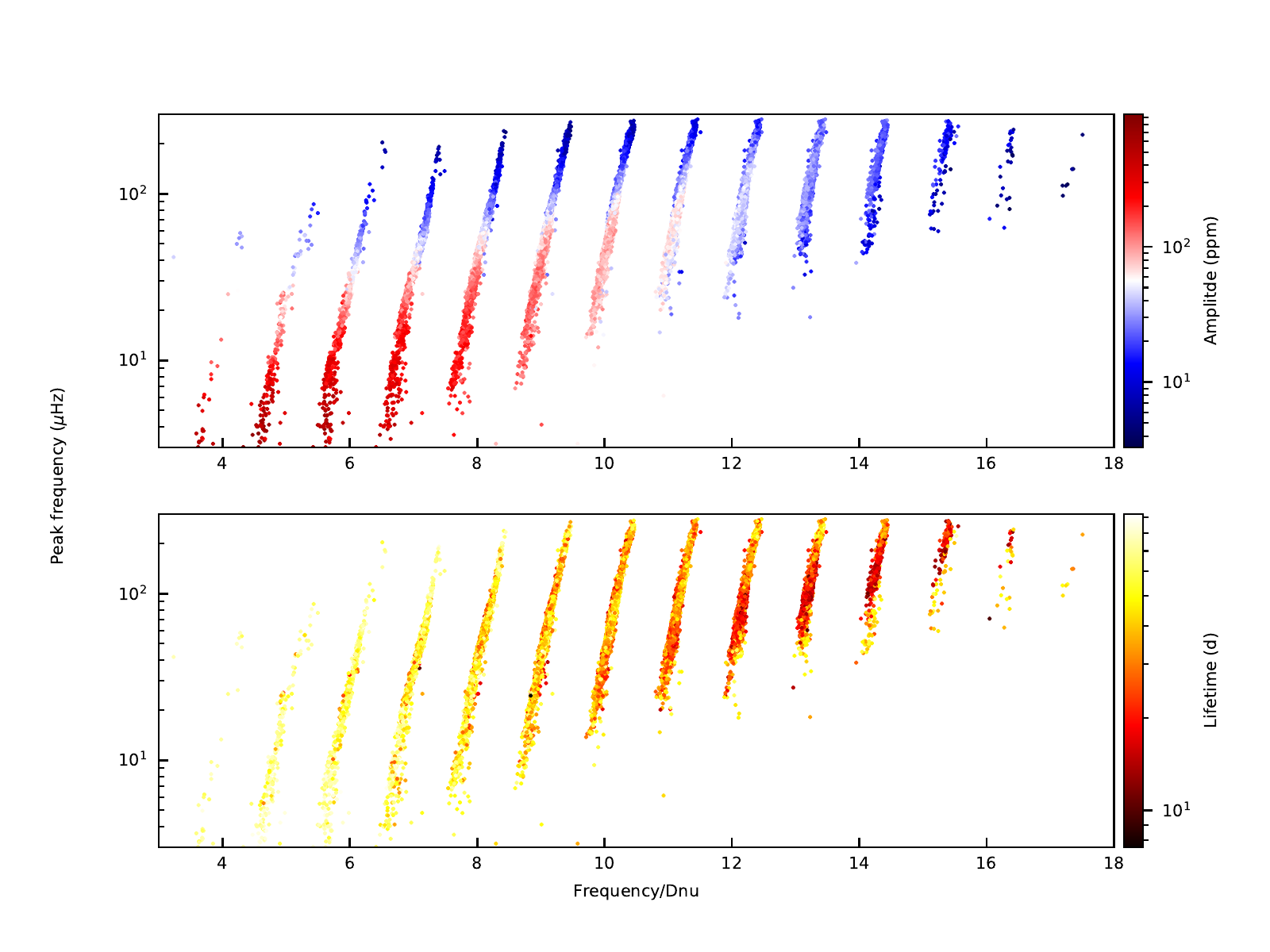}
	\caption{The about 17\,000 radial modes with a mode evidence > 0.99 for 3\,375 H-shell burning stars with color-coded amplitudes (top) and lifetimes (bottom).} 
	\label{fig:radialmodes} 
	\end{center} 
\end{figure}
\begin{figure}[t]
	\begin{center}
	\includegraphics[width=\textwidth]{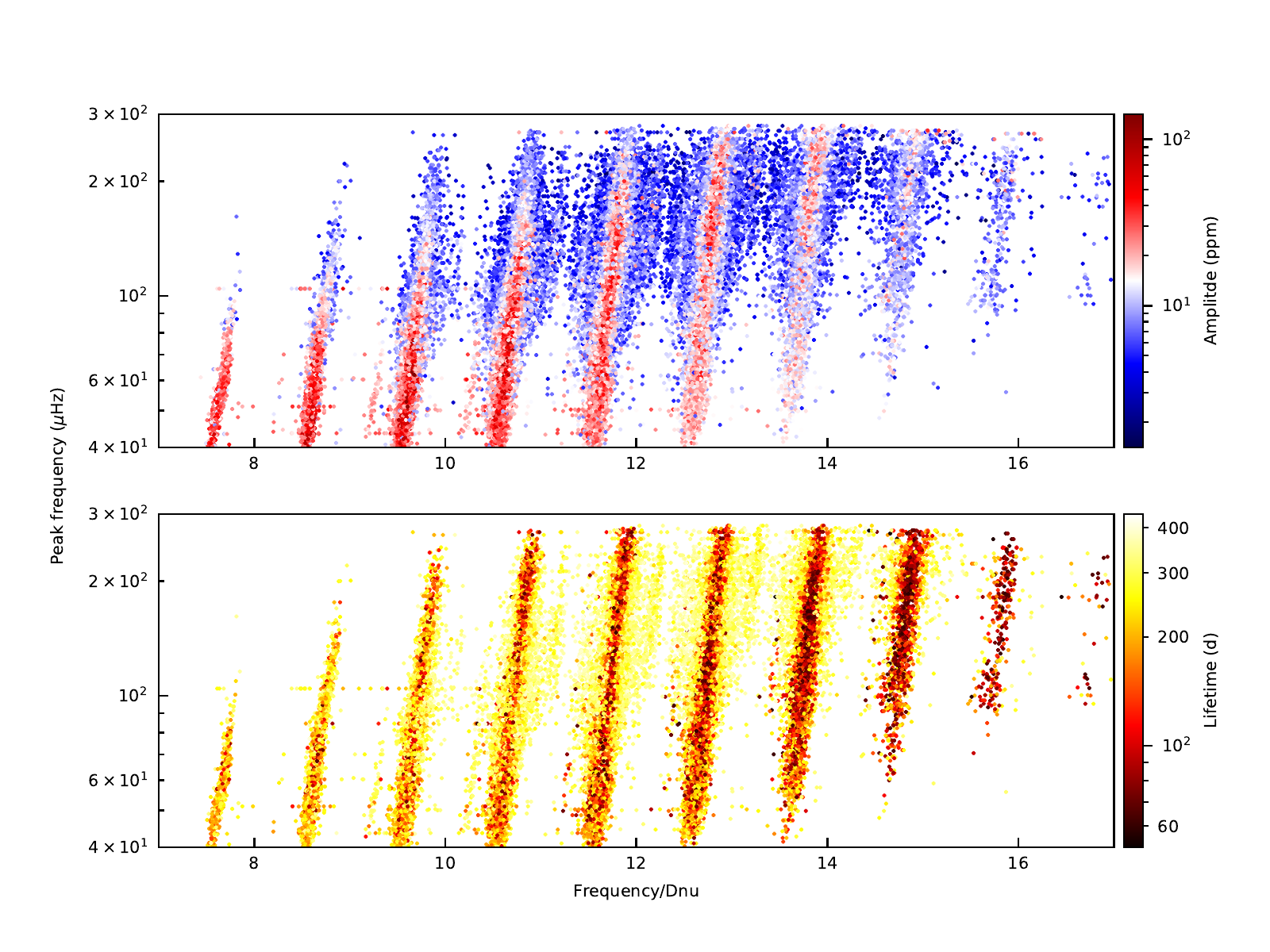}
	\caption{The about 50\,000 $l=1$ modes with a mode evidence > 0.99 for 3\,375 H-shell burning stars with color-coded amplitudes (top) and lifetimes (bottom).} 
	\label{fig:dipolemodes} 
	\end{center} 
\end{figure}

The core of the automated peak-bagging approach is the Bayesian nested sampling algorithm \textsc{MultiNest} \citep{feroz2009,feroz2013}, which performs all fitting tasks within \abba . For the analysis of a single star it is necessary to run \textsc{MultiNest} up to several hundred times. The extensive in- and output is thereby handled by an \textit{IDL} program, which delivers a single file containing all mode parameters and some optional graphical output (available at the Github repository).

\abba\ performs the peak bagging in two steps: 

1) It first fits Lorentzian profiles to all statistically significant \textit{even modes} (i.e. pure pressure $l=0$ and highly p-dominated $l=2$ modes) in the power density spectrum by also considering power from possible $l=1$ modes in the close vicinity of the even modes, which would distort the fits if not taken into account. Examples for such dipole modes are present around 148.5 and 161\mh\ in Fig.\,\ref{fig:rawspec}. Instead of fitting all modes at once, \abba\ handles the even modes per radial order, where their approximate position is estimated from the prerequisite parameters $\nu_\mathrm{c}$, $\Delta\nu_\mathrm{c}$, and $\delta\nu_{02}$. The individual mode significances are evaluated by comparing the statistical evidence of fits based on different scenarios (two modes, only one mode, or no mode is statistically significant). The best-fit model is then ``prewhitened'' in Fourier space from the PDS, which finally contains only power from \textit{odd modes} (i.e. $l=1$ and 3).

2) Using the same principles as above for the odd modes would require detailed knowledge about the dipole mode pattern \citep[e.g.][]{Mosser2012}, which is difficult to obtain. \abba\ follows a different approach, where only a rough estimate of the \textit{local} dipole mode period spacing $\Delta P$ is needed, which is determined from Eq.\,9 of \cite{Mosser2012} based on scalings for the asymptotic period spacing $\Delta\Pi_1$ and the coupling factor $q$. \abba\ then centres a $\Delta P$-wide frequency window on the largest peak in the PDS and fits sequences of one to three (to account for possibly rotationally split modes\footnote{\abba\ does \textit{not} fit rotational splittings but treat all modes and their rotational components as independent. The only connection to ``rotation'' is that it allows for up to three mode profiles in a single local $\Delta P$-wide frequency window, where only one -- potentially rotationally split -- mode should be present.}) Lorentzians and sinc$^2$ (to decide between resolved and unresolved modes) functions to it. Significances are again evaluated by comparing the individual fit evidences delivered by \textsc{MultiNest}. The best-fit model is then prewhitened from the PDS and the procedure starts again (now centred on the new largest peak in the spectrum) until no significant signal is left. Finally, $l=3$ modes are identified from their lifetime and relative position in the radial p-mode order. 

Even though \abba\ fits a single peak in the PDS up to six times, the technique turns out to be much faster than conventional approaches, often fitting many modes at once (with a multitude of free parameters). Extensive tests have shown that it is also quite reliable and especially insensitive to the exact value of $\Delta P$. Even though it frequently happens that the currently analysed window is not properly positioned, this does not really impair the result since missing or additional modes (in the present window) will be picked up in a later step. It is only problematic if the window border is within a mode profile but this barely happens for stars where the dipole pattern is sufficiently well resolved (above a $\nu_\mathrm{max}$ of some ten \mh ).

A detailed description of \abba\ including various tests and comparisons with results from other approaches \cite[e.g.][]{Corsaro2015, Themessl2018} is close to be finished. First results for radial mode amplitudes and lifetimes determined with an older version of \abba\ were presented by \cite{vrard2018}.

Fig.\,\ref{fig:radialmodes} and \ref{fig:dipolemodes} show the radial and $l=1$ modes, respectively, extracted by \abba\ from the 3\,375 RGB stars in the sample, where only modes with an evidence > 0.99 are shown. These plots reveal a lot of interesting features (such as correlations with the evolutionary stage, i.e. $\nu_\mathrm{max}$). A more detailed analysis will be presented in a series of forthcoming papers.

\section{Output files}

An example for the \abba\ output is shown in List.\,1 for the RGB star KIC\,1433803. The modefile contains general information about the star in the header: 
\begin{itemize}
\item {\tt version number}: The current release of the modefiles covers only raw output from \abba , which is of course not perfect. In a later version also some post-processing (e.g. reliability checks of the mode degree, avoid double solutions within the frequency resolution, identify rotational multiplets, etc.) is planned. For now it is the responsibility of the user to check the results for inconsistencies. 
\item {\tt fmax}: The frequency of the maximum oscillation power $\nu_\mathrm{max}$ in \mh\ as defined in Eq.\,2 in \cite{Kallinger2014} with two super-Lorentzian functions with a fixed exponent of four. The $1\sigma$ uncertainty is given as {\tt fmax\_e}.
\item {\tt dnu}, {\tt dnu02}, and {\tt f\_c}: The large and small frequency separation determined in the central three radial orders around $\nu_\mathrm{max}$ and the frequency of the central radial mode as defined in Eq.\,2 of \cite{Kallinger2010a}. All parameters are in \mh .
\item {\tt dnu\_cor} and {\tt alpha}: Curvature corrected large separation in \mh\ and the corresponding dimensionless curvature parameter as defined in Eq.\,4 of \cite{kal2018}.
\item {\tt evo}: Evolutionary state of the star determined from the phase shift of the central radial mode \citep{Kallinger2012} with 0 $\rightarrow$ RGB star, 1 $\rightarrow$ RC star, 2 $\rightarrow$ secondary clump star, and 3 $\rightarrow$ AGB star.
\end{itemize}
The individual mode parameters are given in the main block:
\begin{itemize}
\item {\tt l}: Spherical degree of the mode
\item {\tt freq}: Mode frequency in \mh
\item {\tt amp}: Rms amplitude of the mode in ppm 
\item {\tt tau}: Mode lifetime in days (0 for unresolved modes fitted with a sinc$^2$-function)
\item {\tt ev} and {\tt ev1}: Mode evidence (i.e., the probability that the mode is not due to noise). A useful threshold is 0.91, which corresponds to \textit{strong evidence} in probability theory. The {\tt ev1} parameter is only valid for $l=1$ modes. More details will be given in Kallinger (in prep.).
\end{itemize}

\begin{figure}[t]
	\begin{center}
	\includegraphics[width=\textwidth]{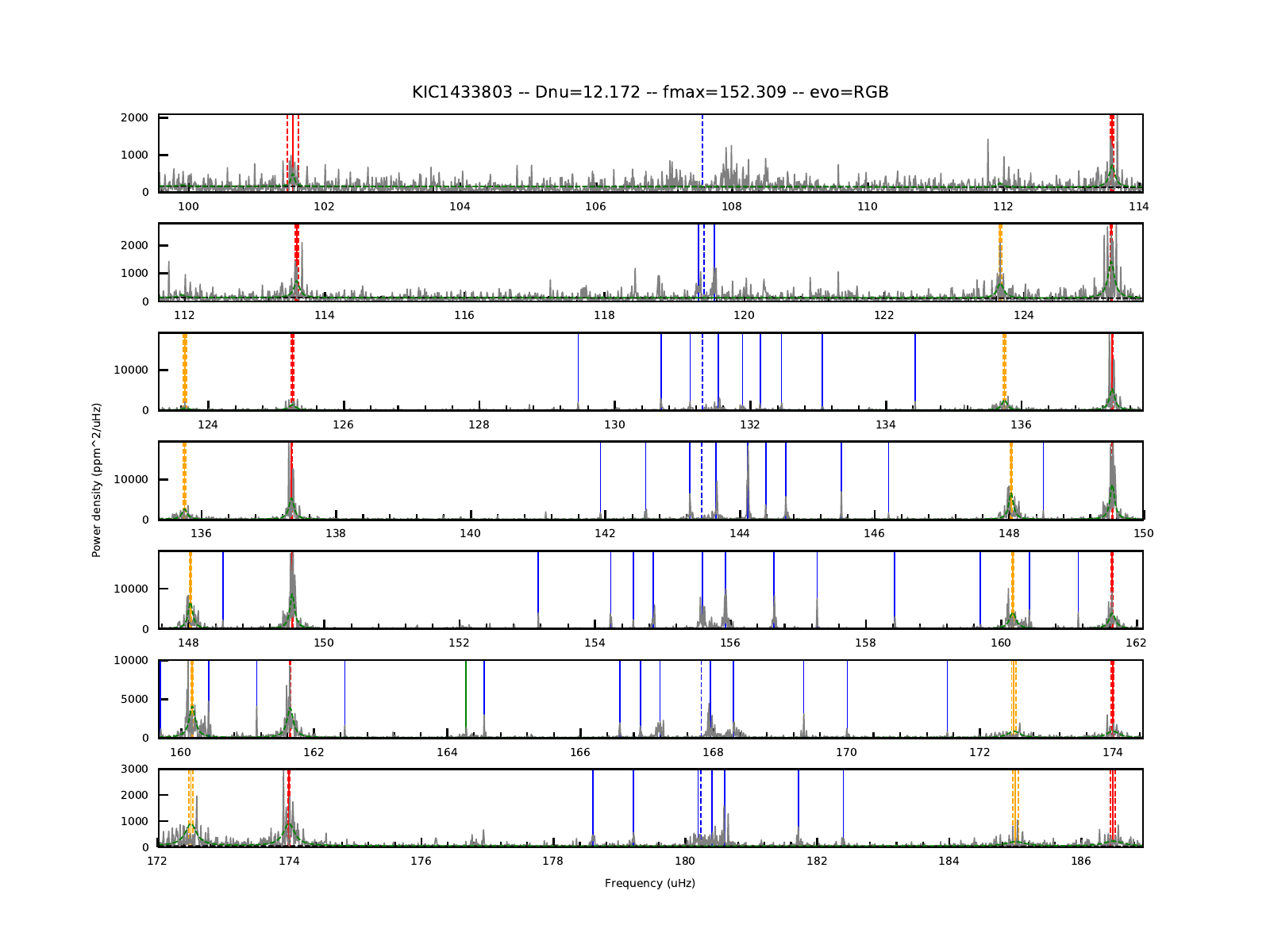}
	\caption{The seven radial p-mode orders of KIC\,1433803 that contain significant oscillation modes. The grey line is the original power density spectrum. The black and green dashed lines show the granulation background signal and a cummulative fit to the $l = 0$ and 2 modes. Vertical red, orange, blue, and green lines indicated the frequencies of the fitted $l =0$, 1, 2, and 3 modes, respectively. For red and orange lines, dashed lines give the $\pm1\sigma$ uncertainties. The blue dashed line gives the assumed postion of the pure pressure dipole mode (i.e., mid-point between two consecutive radial modes).} 
	\label{fig:pds} 
	\end{center} 
\end{figure}

\begin{figure}[t]
	\begin{center}
	\includegraphics[width=\textwidth]{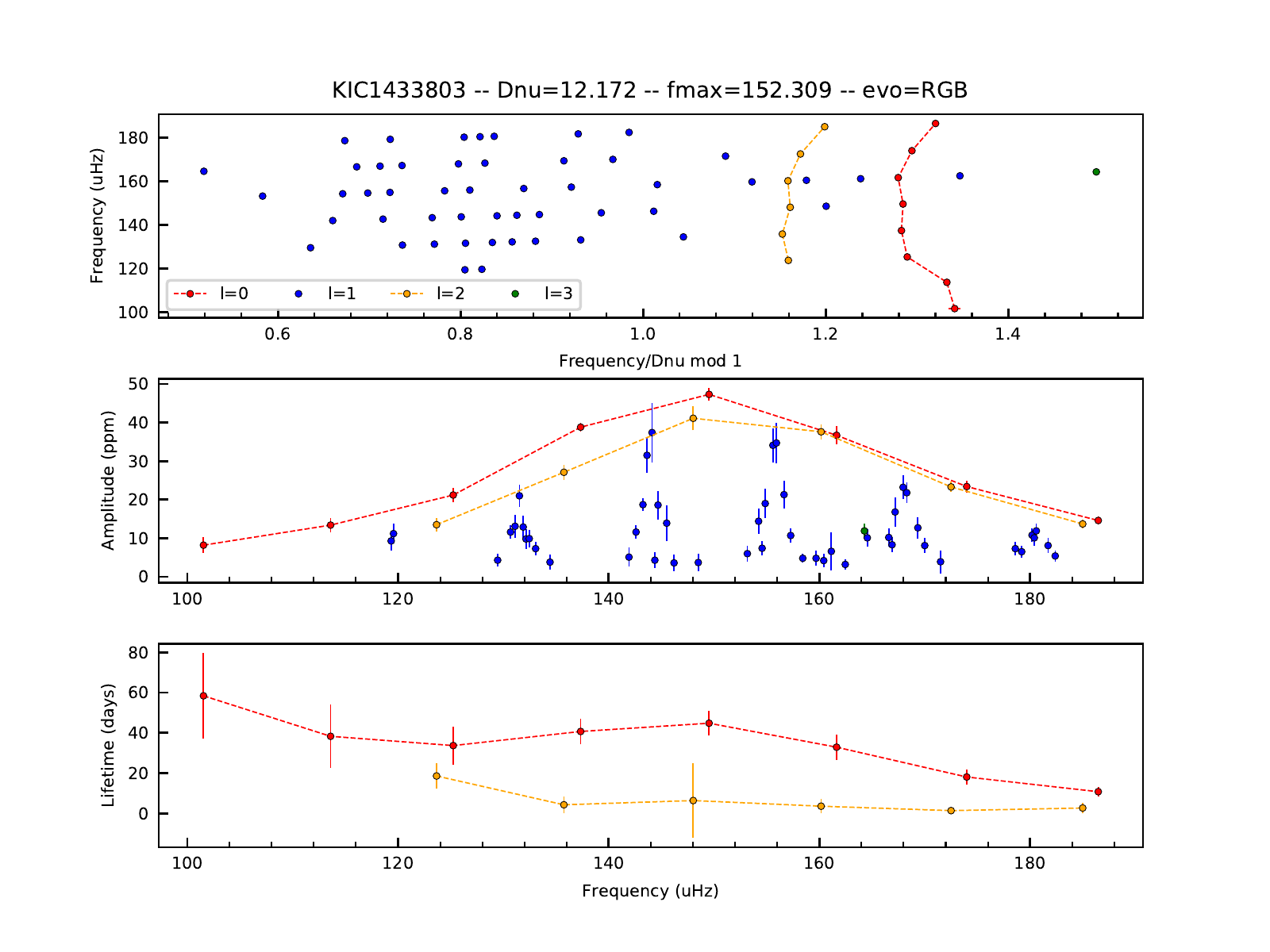}
	\caption{Mode frequencies (top -- in an echelle diagram), amplitudes (middle), and lifetimes (bottom) extracted from the PDS of KIC\,1433803. Frequency uncertainties are also shown in the top panel but are with typically below 20nHz mostly smaller than the symbol size.} 
	\label{fig:echelle} 
	\end{center} 
\end{figure}

The modefiles for the analysed 6\,179 red giants are publicly available at \url{https://github.com/tkallinger/KeplerRGpeakbagging}. Visualisations of the \abba\ output are shown in Fig.\,\ref{fig:pds} and \ref{fig:echelle}, which are also available at the github repository (for all stars). A Python function that loads the modefile of a given star from the github repository into a Pandas dataframe is given in List.\,2.

\begin{lstlisting}[float=*, basicstyle=\ttfamily\tiny, caption= Example for a modefile ({\tt 1433803.modes.dat}) as stored in the Github repository]
version 1.0 (Automatic Bayesian peakBagging Algorithm -- 06/2019 -- T.Kallinger)
#
KIC  1433803
#
      fmax    fmax_e       dnu     dnu_e   dnu_cor dnu_cor_e     dnu02   dnu02_e      f0_c    f0_c_e     alpha   alpha_e   evo
   152.309     0.252    12.172     0.007    12.184     0.004     1.517     0.010   149.504     0.007    0.0034    0.0005     0
#
  l      freq  freq_err       amp   amp_err       tau   tau_err        ev       ev1
  0   101.529     0.081       8.2       2.0      58.4      21.3     0.999     0.000
  0   113.597     0.021      13.4       1.8      38.3      15.7     1.000     0.000
.
.
.
  3   164.278     0.026      11.9       1.9      66.8      29.5     1.000     1.000
\end{lstlisting}

\begin{lstlisting}[language=Python, float=*, basicstyle=\ttfamily\tiny, caption= Python function to load the mode parameters from Github into a Pandas Dataframe.]
import pandas as pd

def read_modefile( kic , header):
	# Python function which loads the modefile of particular RG (define by string <kic>) directly from the 
	# Github repository and returns a Pandas Dataframe containing either the header parameters (header=True) 
	# or the mode parameters (header=None). If the file does not exist None is returned.
	#	Usage:
	#	from read_modefile import read_modefile
	#	freq = read_modefile(<kic>, header=None)	#calling function read_modefile to return mode parameters of <kic>
	#	head = read_modefile(<kic>, header=True)	#calling function read_modefile to return header parameters of <kic>

	url = 'https://raw.githubusercontent.com/tkallinger/KeplerRGpeakbagging/master/ModeFiles/'+kic+'.modes.dat'
	try :
		if header :
			data = pd.read_csv(url, delimiter=' ', skiprows = 5, nrows=1, skipinitialspace = True, header = None, 
				names = ['fmax','fmax_e','dnu','dnu_e','dnu_cor','dnu_cor_e','dnu02','dnu02_e','f0_c','f0_c_e','alpha','alpha_e','evo'] )
		else :
			data = pd.read_csv(url, delimiter=' ', skiprows = 8, skipinitialspace = True, header = None, 
				names = ['degr', 'freq', 'freq_e', 'amp', 'amp_e', 'tau', 'tau_e', 'ev', 'ev1'] )
	except :
		data = None
	return data
\end{lstlisting}

\section*{Acknowledgements}
\textit{I would like to thank Paul G. Beck for many fruitful discussions and inspirations. I am also grateful for funding via the Austrian Space Application Programme (ASAP) of the Austrian Research Promotion Agency (FFG) and acknowledge the \textit{Kepler Science Team} and all those who have contributed to making the \textit{Kepler} mission possible. Funding for the \textit{Kepler Discovery mission} is provided by NASA's Science Mission Directorate.} 

\bibliography{peakbag}{}
\bibliographystyle{apalike}


\end{document}